# Anharmonic Coupling between Intermolecular Motions of Water Revealed by Terahertz Kerr Effect

Tobias Kampfrath, Martin Wolf and Mohsen Sajadi*
*Fritz-Haber-Institut der Max-Planck-Gesellschaft, Berlin, Germany.*

*sajadi@fhi-berlin.mpg.de

**Formation of local molecular structures in liquid water is believed to have marked effect on the bulk properties of water, however, resolving such structural motives in an experiment is challenging. This challenge might be handled if the relevant low-frequency structural motion of the liquid is directly driven with an intense electromagnetic pulse. Here, we resonantly excite diffusive reorientational motions in water with intense terahertz pulses and measure the resulting transient optical birefringence. The observed response is shown to arise from a particular configuration, namely the restricted trans-lational motion of water molecules whose motions are predominantly orthogonal to the dipole moment of the excited neighboring water molecules. Accordingly, we estimate the strength of the anharmonic coupling between the rotational and the restricted translational degrees of freedom of water.**

Knowledge about the microscopic structural conformations in liquid water is essential for understanding its function in chemical and biological processes[1,2]. As water molecules are transiently linked to their neighbors with flexible fluctuating hydrogen bonds, the local structures formed in water comprise of conformations with different densities[3,4]. Spectroscopic techniques to unravel this complexity have limitations: for example, dielectric relaxation and Raman spectroscopy measure only the time- and ensemble- averaged dynamics by which the local structural information is obscured. On the other hand, although time-resolved multi-dimensional spectroscopies such as photon-echo[5] can in principle resolve the structural heterogeneities[6,7], their implementation in water is challenging[8]. Fortunately, recent advances based on the resonant excitation of the low-frequency structural modes of liquids have provided an alternative path to study the microscopic structural dynamics of liquids[9].

In line with the latter approach, we resonantly excite rotational degrees of freedom of water with intense terahertz (THz) pulses. The resulting anisotropy is monitored by an optical probe pulse in a THz Kerr effect (TKE) configuration[10,11]. The transient birefringence of water relaxes with a time constant of ~0.5 ps and, compared to optical excitation, exhibits strongly enhanced amplitude of opposite sign. We assign the TKE response to the restricted translational motion of water molecules and suggest a picture of intermolecular anharmonic coupling between the excited rotational motion and the restricted translational motion of the neighboring water molecules. We show the observed response originates from a conformation in which the latter two modes are predominantly orthogonal.

**Experimental**. A schematic of the TKE experiment is shown in **Fig. 1a**. An intense linearly polarized THz electric field (peak strength of approximately 2 MV/cm) is used to excite the water sample. The resulting optical birefringence $\Delta n(t)$ is measured by a probe pulse (800 nm, 2 nJ, 8 fs) whose linear polarization acquires ellipticity when traversing the sample. The distilled liquid water film (thickness of 50 μm) is held between a rear glass window and a 150 nm thick silicon nitride membrane as entrance window whose Kerr response makes a negligible contribution to the TKE signal[12].

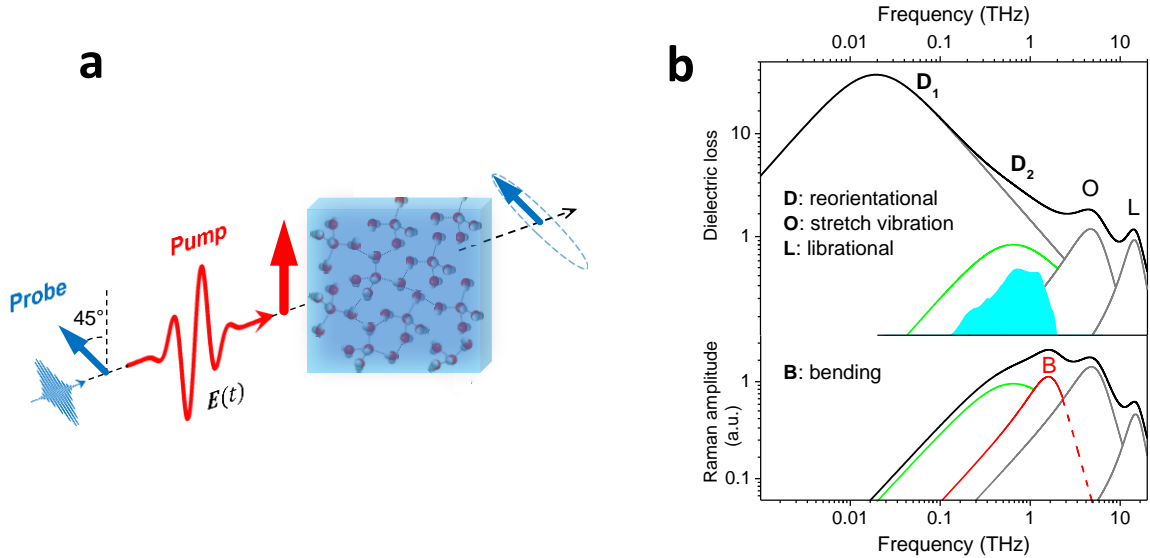

**Fig. 1. Dynamic THz Kerr effect. a**, An intense THz pump pulse is used to induce optical birefringence in water, which is monitored by an optical probe pulse that becomes elliptically polarized upon traversing through the medium. **b**, Equilibrium dielectric loss ($\varepsilon''$) and Raman spectra of water[21]. The spectrum of the excitation THz field is indicated by the cyan area.

The THz-field-induced optical birefringence of water is compared to that induced by an optical pump pulse (Optical Kerr effect, OKE). Both optical and THz excitation are conducted in the same setup under otherwise identical conditions. The intensity envelopes of the optical and the THz pump pulses have approximately the same temporal width of about 350 fs (see **Fig. 2a** and **2b**), thereby allowing for straightforward comparison of the TKE and the OKE data.

**Results**. The transient birefringence of water following optical excitation is shown in **Fig. 2a**. The unipolar shape of the signal is in line with all previously reported OKE signals of water[13,14,15,16,17]: a signal with a spike around time zero (due to the instantaneous electronic response) and a relatively weak relaxation tail extending over few picoseconds (due to the nuclear response of the liquid). This behavior is in strong contrast to the birefringence induced by the THz pulse shown in **Fig. 2b**. In particular, the TKE signal of water has a bipolar shape. Note that the TKE signal does not change when the THz field is reversed (see green and blue curves in **Fig. 2b**). This behavior is in line with the quadratic dependence of the TKE signal on the THz field amplitude (see **Fig. 2c**).

The experimental results of **Fig. 2a** and **2b** reveal three distinct differences between the TKE and the OKE signals of water. (i) In stark contrast to its OKE signal and also the TKE signals of simple liquids[18], the TKE response of water is bipolar. Compared to the OKE signal of water, the relaxation tail of the TKE signal has reversed (negative) sign, also relative to the instantaneous electronic spike around time zero.

(ii) In comparison to the OKE signal of water, the amplitude of the nuclear part of the TKE signal is enhanced by a factor of ~30. The enhancement value is obtained by normalizing both the signals to the peaks of their instantaneous electronic responses. This approach is identical to normalization of the signal to the pump intensity, as both pump pulses have comparable duration[18]. Note that in **Fig. 2** the TKE signal is shown on the linear scale whereas the OKE signal is on a logarithmic scale.

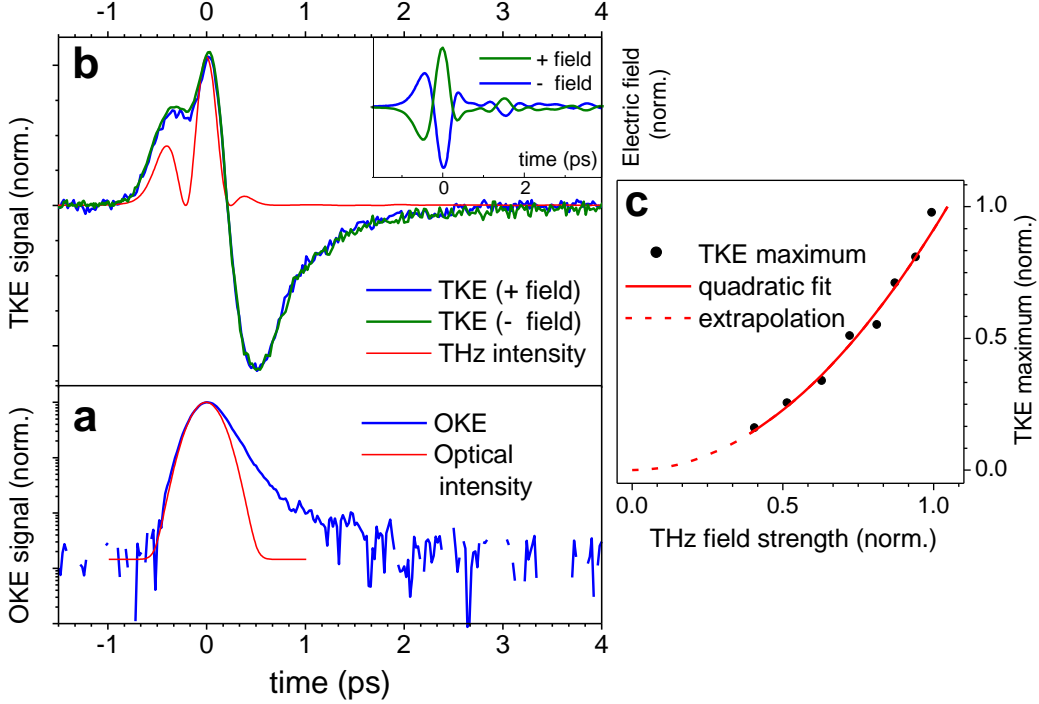

**Fig. 2. Dynamic Kerr effect of water. a**, Optical Kerr effect of water at 22°C (blue line). Red line shows the intensity of optical pump pulse estimated as a Gaussian fit to the left flank of OKE response of water. Note that the OKE signal is presented in logarithmic scale. **b**, THz-field-induced transient optical birefringence of water (blue and green lines) and the square of the driving THz field (red line). THz electric fields with opposite polarities (see inset) result the same TKE signal of water. **c**, Fluence dependence shows that the TKE of water scales quadratically with THz driving field.

(iii) The tail of the TKE signal of water relaxes with a time constant of 0.5 ps. Because of its enhanced amplitude, we can determine this relaxation time with high accuracy. In contrast, the smaller magnitude of the OKE signal only allows us to extract one relaxation component with a time constant of about 0.5 ps. However, time constants exceeding 1ps of the OKE response of water have been reported previously by using more specialized OKE setups[13,14,15,16,17].

**Discussion**. In general, the dynamic Kerr signal can be decomposed into (i) an electronic and (ii) a nuclear contributions. The latter contribution may be expressed by the following correlations[19]

$$R^{\text{nuc}} = \langle[\Pi(t_1), \Pi(0)]\rangle + \langle[\Pi(t_1 + t_2), [\mu(t_1), \mu(0)]]\rangle. \qquad (1)$$

Here, $\Pi$ and $\mu$ refer respectively, to the total polarizability anisotropy and the permanent dipole moment of the liquid, $[A, B] = AB - BA$, the angle bracket denotes an ensemble average. The total polarizability $\Pi$ encompasses the polarizability of single molecules, $\Delta\alpha$ and also the induced polarizability $\Pi^I$ via intermolecular interactions/collisions. The auto-correlation term in Eq. (1) illustrates a pure polarizability relaxation, while the second term contains the cross correlation of the polarizability tensor and the permanent dipole moment.

*Deconvolution.* To unravel the underlying dynamical components of the TKE signal of water, we reproduce the TKE signal by a convolution of the driving THz intensity with exponential functions. As illustrated in **Fig. 3**, the square of the THz pulse, taken from the TKE signal of a thin diamond plate (black line) is convoluted by two decaying exponential functions whose time constants and relative amplitude are left as fitting parameters. The addition of the convoluted components with time constants of 0.1 ps (cyan line) and 0.5 ps (red line) reproduces the experimental results (green line in **Fig. 3**), reasonably well. Therefore, the TKE response of water can be understood as an instantaneous electronic response around time zero and two step-like nuclear responses with opposite sign that decay exponentially with time constants of about 0.1 ps and 0.5 ps, respectively. In the following, we take use these distinct features of the TKE signal to develop a consistent interpretation of the signal.

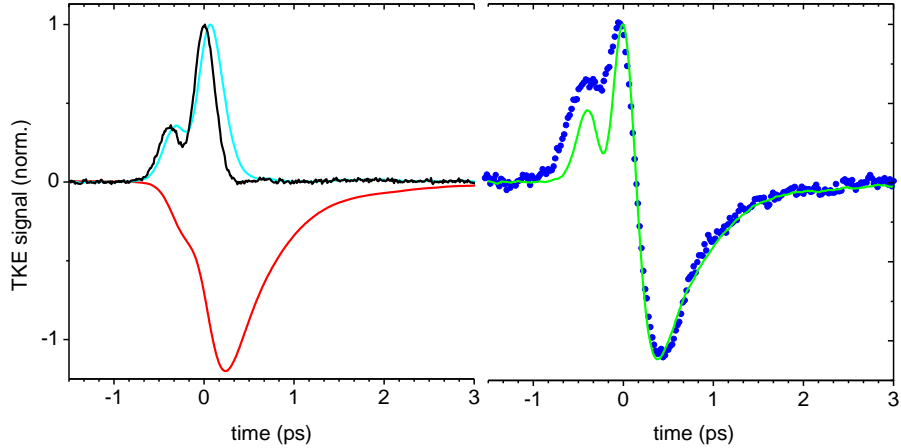

**Fig. 3. Dynamic components in THz Kerr effect signal of water. left**, Square of THz field (black line) is convoluted with two exponential functions with $\tau_1 \approx 0.1$ ps (cyan line) and $\tau_2 \approx 0.5$ ps (red line). **right**, Subtraction of the latter two curves gives rise to the green line which captures almost all features of the TKE signal of water (blue dots).

*Amplitude enhancement.* The dielectric spectrum of water up to 30 THz is shown in **Fig. 1b** and is typically described by two relaxation processes with time constants of 8.4 ps and 0.4 ps plus two vibrations with resonance frequencies of 5 THz and 15 THz[20,21]. The bending of H-bonds at ~2 THz is strongly Raman active but has a minor contribution in the dielectric spectrum of water[22,23,20]. Our THz excitation pulse (cyan area) is in resonance with reorientational diffusive motions in water ($D_1$ and $D_2$ processes). As a result, the pump field **E** exerts the torque $\boldsymbol{\mu} \times \mathbf{E}(t)$ on the permanent dipole moment $\boldsymbol{\mu}$ [11,18]. This torque gives rise to a larger transient birefringence in the liquid than the torque $\boldsymbol{\mu}_{\text{ind}} \times \mathbf{E}(t)$ applied to the induced dipole moment $\mu_{\text{ind}} = \Delta\alpha E$. Therefore, for water as a polar liquid with isotropic polarizability[24], the second term in Eq. (1) provides the dominant contribution to the THz-field induced anisotropy in the liquid.

Despite this assignment, the microscopic molecular origin of the enhanced negative component of the TKE signal needs to be determined. It may arise from a single molecule reorientational relaxation or alternatively from a different intermolecular mode(s). In the latter scenario, the resonantly excited motion would act as an intermediate state to receive the external THz energy and instantaneously transfer it to other modes. As discussed in the next sections, the time constant and the sign of the relaxation tail of the TKE signal are the main elements to decide between these two scenarios.

*Reverse sign.* A distinct feature of the TKE signal of water compared to many other liquids (more than 20 liquids studied by us including *n*-alcohols) is the bipolarity of the signal. As the TKE response of polar molecules scales linearly with $\Delta\alpha$, the sign of the TKE signal obeys the sign of $\Delta\alpha$[25,26]. This correlation was recently corroborated for molecular gases. We showed that the TKE response of fluoroform ($CHF_3$) with $\Delta\alpha < 0$ has a reverse sign compared to the TKE signal of acetonitrile vapor with $\Delta\alpha > 0$[26], and in the case of water, our gas-phase TKE experiment clearly declares a positive sign for the polarizability anisotropy of water i.e. $\Delta\alpha > 0$[26]. Therefore, it is unlikely that the TKE response of water originates from the single molecule reorientational relaxation.

Note that in contrast to the simple aprotic liquids, the Debye processes in water may even encompass the reorientational motion of water clusters[27]. Therefore, the observed TKE response of water may arise from a cluster structure whose sign of polarizability anisotropy is negative. This scenario is also supported by our temperature dependent TKE signal of water (see **Fig. S1**). Here, we observe an increase of the signal amplitude by decreasing the temperature, in concomitance with the increase of the fraction of H-bonded molecules at lower temperature[28]. In the following we suggest a simple molecular conformation, by which the coupling of the THz electric field to a permanent dipole moment and a negative sign for its polarizability anisotropy can be explained.

As the TKE signal of CHF$_3$ clearly shows a sign reverse[26], we use the molecular structure of CHF$_3$ to find a similar structure in water. The permanent dipole moment of CHF$_3$ along its C-H bond is less polarizable than the C-F bonds (see **Fig. 4**), resulting in a negative polarizability anisotropy of this molecule, $\Delta\alpha = \alpha_\| - \alpha_\perp < 0$[29,30]. For a symmetric top molecule such as CHF$_3$, the $\alpha_\|$ and $\alpha_\perp$ are the polarizability anisotropy components parallel and perpendicular to the molecular symmetry axis. A similar arrangement of a permanent dipole and a polarizable mode may also be realized in the simplest cluster of water, i.e a water dimer. However, here the polarizability arises from an intermolecular interaction/collision among molecules. Indeed, as shown in **Fig. 4** one may find a supramolecular type structure formed by two adjacent water molecules wherein the permanent dipole moment of the first water molecule and the restricted translational motion of its neighbor are predominantly orthogonal. In the structure highlighted in **Fig. 4**, the O-H group provides the permanent dipole moment of the supramolecule, whereas the restricted translational motion of the second molecule provides the perpendicular component of its polarizability anisotropy. The latter translational motion causes the bending of the H-bonds and is strongly Raman active (B mode in **Fig. 1b**)[31]. Analogous to the CHF$_3$, the polarizability anisotropy of this supramolecule is negative $\Pi^I = \Pi^I_\| - \Pi^I_\perp < 0$, as the polarizability of the bending mode along the O-H group is smaller than its perpendicular component. As a result, the TKE response, $\langle[\Pi^I(t_1+t_2),[\mu(t_1),\mu(0)]]\rangle$ of this supramolecule gains a negative sign. Moreover, because of the contribution of the permanent dipole moment $\mu$ in the response, the measured signal is enhanced compared to the OKE response, $\langle[\Pi^I(\tau),\Pi^I(0)]\rangle$. Finding larger clusters with analogous properties should be examined in molecular simulations.

*Relaxation time constant.* Our assignment of the negative nuclear component of the TKE response of water to the restricted translational motion of water molecules can also be endorsed by its relaxation time constant. In the Raman spectrum of water the latter mode is typically fit by a Lorentzian function with resonance frequency and damping rate of $\Omega_B/2\pi \approx 50$ cm$^{-1}$, $\gamma_B \approx 115$ cm$^{-1}$, respectively[21,28]. Therefore for such a damped motion, the dynamics may be resolved as an exponential relaxation $e^{-\frac{\gamma_B}{2}t}$ with $\tau = (\frac{\gamma_B}{2})^{-1} \approx 0.5$ ps. Interestingly, this time constant agrees well with the one resolved in our TKE experiment. On the other hand, because of the weak polarizability of water, the reported OKE responses are diverse and their interpretations have been controversial[13,14,15,16,17]. However, the assignment of an intermediate time constant of about $\tau \approx 0.5$ ps to the restricted translational motion of water has also been suggested[32,33].

**Model**. We suggest the following molecular mechanism for the TKE response of water. The THz driving field applies a torque to the permanent dipole moment of the H-bonded water molecules. The momentum resulting from this torque is transferred instantaneously to the restricted translational motion of a neighboring water molecule via an anharmonic coupling between these two intermolecular degrees of freedom. Eventually, the induced anisotropy in the restricted translational mode relaxes with a time constant of ~0.5 ps. In this picture, the interaction energy may be given by $V_1 = -\frac{g}{2}Q_B Q_\theta^2$, where $g$ is the anharmonic coupling constant and $Q_\theta$ ($Q_B$) stands for the rotational (restricted translational) coordinate. Under this coupling potential the equation of motion of the $Q_B$ mode is expressed by

$$\ddot{Q}_B + \gamma_B \dot{Q}_B + \Omega_B^2 Q_B = -\frac{\partial}{\partial Q_B}V(t) = \frac{g}{2M}Q_\theta(t)^2 \qquad (2)$$

where $\Omega_B = 2\pi\nu_B$, $M$ is the mass of a water molecule and $Q_\theta(t)$ illustrates the dynamics of a single molecule rotational coordinate in which the THz field couples in. We obtain the dynamics of $Q_\theta(t)$ from the diffusion equation which is driven by $\boldsymbol{\mu}_0 \times \mathbf{E}_{THz}(t)$[18]. As the response should scale quadratically relative to $\mathbf{E}_{THz}(t)$ (see **Fig. 2c**), we seek a solution of the diffusion equation which is proportional to $Q_\theta(t)^2$ (as the alignment of molecules by transient external electric fields is small, we assume the alignment factor $\langle\cos^2\theta(t)\rangle$ is directly proportional to $Q_\theta(t)^2$). The final result is then given by[18]

$$Q_\theta(t)^2 = 2\beta^2\mu_0^2 R_2(t) * \left[E_{THz}(t) \cdot \mathcal{F}^{-1}\left(\frac{E_{THz}(\omega)}{1+i\omega\tau_D}\right)\right]. \qquad (3)$$

Here $\mathcal{F}^{-1}$ and $*$ stand respectively, for the inverse Fourier transform and convolution operators and $\beta = 1/K_B T$. The decay of the resulting change in the angular distribution of water molecules may then be captured by the response function $R_2(t) = e^{-3t/\tau_{D2}}$ with $\tau_{D2} \approx 0.4$ ps, where $\tau_{D2}$ refers to the time constant of the fastest Debye process of water. Therefore, $Q_\theta(t)^2$ relaxes with a time constant of about 100 fs.

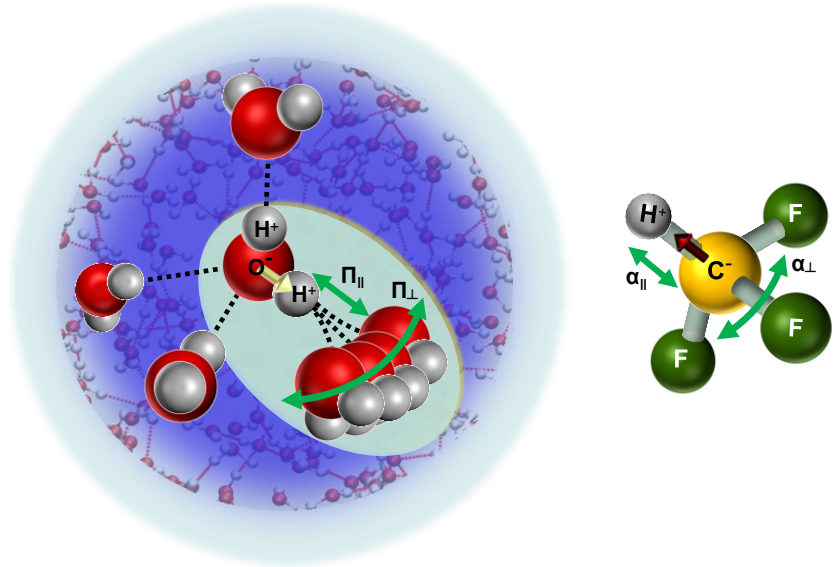

**Figure 4 | Intermolecular mode coupling in water. right**, In fluoroform the C-H bond along the symmetry axis of the molecule is less polarizable than the C-F bonds, thereby the polarizability anisotropy of this molecule is negative, $\Delta\alpha = \alpha_\parallel - \alpha_\perp < 0$. The sign of $\Delta\alpha$ is directly imprinted on the sign of TKE response of this molecule (see main text and Ref. 26). **left**, In water a supramolecular structure with a permanent dipole (O-H bond) and a component of polarizability anisotropy perpendicular to the O-H bond -similar to fluoroform- can be envisaged. In this unit cell, the total polarizability anisotropy is negative $\Pi^I = \Pi^I_\parallel - \Pi^I_\perp < 0$, because $\Pi^I_\parallel$ corresponds to the restricted translational motion of a single water molecule. Analogous to fluoroform, one may expect a negative TKE response from this supramolecule.

As $Q_\theta(t)^2$ decays very fast, we approximate its dynamics with a delta function; hence the driving force in Eq. 2 simplifies to $M^{-1}\beta^2\mu_0^2 E_{THz}^2 \delta(t)$. The $Q_B$ coordinate can also be triggered directly by an optical pulse in a Raman process. In this case, the driven force is $\frac{1}{2M}\frac{\partial\Pi^I}{\partial Q_B}E_{opt}^2\delta(t)$[34].

To estimate the strength of the anharmonic coupling $g$, we compare the quotient of the induced birefringence via THz and optical pulses, $\frac{\Delta n_{\text{THz}}}{\Delta n_{\text{OKE}}}$ with the experimental result. Using the relations $\Delta n = \frac{2\pi}{n}\frac{\partial \Pi^I}{\partial Q_B}Q_B(t)$[35] and $\frac{\Delta n_{\text{THz}}}{\Delta n_{\text{OKE}}} \propto \frac{Q_{B_{\text{THz}}}(t)}{Q_{B_{\text{opt}}}(t)} = \frac{g\beta^2\mu_0^2}{\frac{1}{2}\frac{\partial \Pi^I}{\partial Q_B}} \approx 30$ we obtain

$$g \approx 15\, \beta^{-2}\mu_0^{-2}\frac{\partial \Pi^I}{\partial Q_B} \approx 1 \times 10^{-8}\ \text{Jm}^{-1}. \qquad (4)$$

Here, we assumed T= 295 K, $\mu_0$= 1.5 D, $\frac{\partial \Pi^I}{\partial Q_B} \approx 1.5 \times 10^{-27}$ m$^2$ and $E_{\text{opt}}^2 = E_{\text{THz}}^2$. This coupling constant can be used to determine the anharmonic potential $V_1$ for the maximum displacements in the coordinates $Q_B$ and $Q_\theta^2$. Consequently, $V_1$ contribution to the total potential energy is obtained being small relative to $V_0 = \frac{1}{2}M\Omega_B^2 Q_B^2$. (see supplementary material for details). As a result, the rotational motion of water molecules provides a small anharmonicity to the bending mode of water.

In conclusion, we showed that the measured TKE signal of water arises from the restricted translational motions of molecules whose motions are predominantly orthogonal to the O-H dipole of their neighboring molecules. This picture suggests that our measurement is selective to one particular conformation in the liquid. This notion is supported by the temperature dependent data, as the fraction of four-coordinated tetrahedral conformation is expected to increase by decreasing temperature. This approach opens a new avenue to use an intermolecular mode of water as a local probe of the structural deformation of the H-bonding network of water in the vicinity of solutes, such as ions and biomolecules.

**Acknowledgment**. We thank Takaaki Sato and Richard Buchner for the dielectric fit parameters of water and Kramer Campen for discussions.

**References**

1. D. Eisenberg and W. Kauzmann, The Structure and Properties of Water (Oxford University Press, New York, 1969).
2. G. E. Walrafen, in Water: A Comprehensive Treatise, edited by F. Franks (Plenum, New York, 1972).
3. S. Woutersen, U. Emmerichs, H. J. Bakker, *Science* **278**, 658-660 (1997).
4. F. Rao, S. Garrett-Roe, P. Hamm, *J. Phys. Chem. B,* **114**, 15598-15604 (2010).
5. A. Tokmakoff, M. D. Fayer, *J. Chem. Phys.* **103**, 2810-2826 (1995).
6. Milne, C. J.; Li, Y. L.; Miller, R. J. D. Torre, R., Ed.; (Springer-Verlag: New York, 2008; pp 1-72)
7. Y. Tanimura, A. Ishizaki, *Acc. Chem. Res.* **42**, 1270 (2009).
8. S. Palese et *al., J. Phys. Chem.* **98**, 12466-12470 (1994).
9. J. Savolainen, S. Ahmed, P. Hamm, Pro*c. Natl. Acad. Sci. U. S. A.* **110**, 20402-20407 (2013).
10. M. C. Hoffmann, N. C. Brandt, H. Y. Hwang, K. L. Yeh, K. A. Nelson, *Appl. Phys. Lett.* **95**, 231105-3 (2009).
11. S. Fleischer, Y. Zhou, R. W. Field, K. A. Nelson, *Phys. Rev. lett.* **107**, 163603-5 (2011).
12. M. Sajadi, M. Wolf, T. Kampfrath, *Optics express* **23**, 28985-28992 (2015).
13. E. W. Castner, Y. J. Chang, Y. C. Chu, G. E. Walrafen, *J. Chem. Phys.* **102,** 653-659 (1995).
14. S. Palese, S. Mukamel, R. J. D. Miller, W. T. Lotshaw, *J. Phys. Chem.* **100,** 10380-10388 (1996).
15. S. Palese, L. Schilling, R. J. D. Miller, P. R. Staver, W. T. Lotshaw, *J. Phys. Chem.* **98**, 6308-6316 (1994).
16. Y. J. Chang, E. W. Castner, *J. Chem. Phys.* **99**, **7**289-7299 (1993).
17. R. Torre, P. Bartolini, R. Righini, Nat*ure* **428,** 296-299 (2004).
18. M. Sajadi, M. Wolf, T. Kampfrath, *Nat. Commun.* **8,**15796 (2017).
19. R. W. Hellwarth, *Prog. Quant. Electron.* **5***,* 1-68 (1977).
20. H. Yada, M. Nagai, K. Tanaka, *Chem. Phys. Lett.* **464**, 166-170 (2008).
21. T. Fukasawa et al*., Phys. Rrev. Lett.* **95**, 197802-4 (2005).

22. P. L. Silvestrelli, M. Bernasconi, M. Parrinello, *Chem. Phys. Lett.* **277**, 478-482 (1997).
23. J. B. Hasted, S. K. Husain, F. A. M. Frescura, J. R. Birch, *Chem. Phys. Lett.* **118**, 622-625 (1985).
24. W. F. Murphy,. *J. Chem. Phys.* **67**, 5877-5882 (1977).
25. I. A. Finneran, R. Welsch, M. A. Allodi, T. F. Miller, G. A. Blake, *Proc. Natl. Acad. Sci. U. S. A.* **113**, 6857-6861 (2016).
26. T. Kampfrath, M. Wolf, M. Sajadi, http://arxiv.org/pdf/1706.09623 (2017).
27. I. Popov, P. Ben Ishai, A. Khamzin, Y. Feldman, *Phys. Chem. Chem. Phys.* **18**, 13941-13953 (2016).
28. S. Krishnamurthy, R. Bansil, J. Wiafeakenten, *J. Chem. Phys.* **79**, 5863-5870 (1983).
29. C. K. Miller, B. J. Orr, J. F. Ward, *J. Chem. Phys.* **74**, 4858-4871 (1981).
30. R. Kobayashi, R. D. Amos, H. Koch, P. Jorgensen, *Chem. Phys. Lett.* **253**, 373-376 (1996).
31. G. E. Walrafen, M. R. Fisher, M. S. Hokmabadi, W. H. Yang, *J. Chem. Phys.* **85**, 6970-6982 (1986).
32. K. Winkler, J. Lindner, H. Bursing, P. Vohringer, *J. Chem. Phys.* **113**, 4674-4682 (2000).
33. K. Winkler, J. Lindner, P. Vohringer, *Phys. Chem. Chem. Phys.* **4**, 2144-2155 (2002).
34. M. D. Levenson, S. S. Kano, Introduction to Nonlinear Laser Spectroscopy (Academic Press, San Diego, 1988).
35. R. Merlin, *Solid State Commun.* **102**, 207-220 (1997).